\documentclass[11pt]{article}

\usepackage{setspace} 

\usepackage{amsmath}

\usepackage{graphicx}

\usepackage[round,numbers,sort&compress]{natbib} 

\title{Dynamic instability of a growing adsorbed polymorphic filament}

\author{{\bf Stefano Zapperi}\\
Consiglio Nazionale delle Ricerche- \\ Istituto per l'Energetica e le Interfasi,\\
Via. R. Cozzi 53, 20125 Milano, Italy and \\
Institute for Scientific Interchange Foundation, \\ 
Viale San Severo 65, 10133 Torino, Italy
	\and  {\bf L. Mahadevan} \thanks{Corresponding author: 
School of Engineering and Applied Sciences, 
Harvard University Cambridge, MA 02138, USA. Tel. 617-496-9599 }\\ 
School of Engineering and Applied Sciences, \\
Harvard University Cambridge, MA 02138 \\
and Department of Systems Biology, \\ 
Harvard Medical School, Boston, MA 02115, USA
}

\date{}

\begin{document}

\maketitle

\abstract{The intermittent transition between slow growth and rapid shrinkage in polymeric assemblies is termed dynamic instability, a feature  observed  in a variety of biochemically distinct assemblies including microtubules, actin and their bacterial analogs. The existence of this labile phase of a polymer has many functional consequences in cytoskeletal dynamics, and its repeated appearance suggests that it is relatively easy to evolve. Here, we consider the minimal ingredients for the existence of dynamic instability by considering a single polymorphic filament that grows by binding to a substrate, undergoes a conformation change, and may unbind as a consequence of the residual strains induced by this change. We identify two parameters that control the phase space of possibilities for the filament: a structural mechanical parameter that characterizes the ratio of the bond strengths along the filament to those with the substrate (or equivalently the ratio of longitudinal to lateral interactions in an assembly), and a kinetic parameter that characterizes the ratio of time scales for growth and conformation change. In the deterministic limit, these parameters serve to demarcate a region of uninterrupted growth from that of collapse. However, in the presence of disorder in either the structural or the kinetic parameter the growth and collapse phases can coexist where the filament can grow slowly, shrink rapidly, and transition between these phases, thus exhibiting dynamic instability. We exhibit the window for the existence of dynamic instability in a phase diagram that allows us to quantify the evolvability of this labile phase.

}

\section*{Introduction}

Polymeric filaments are the building blocks of nearly all biological structures at the cellular level. Their structural, chemical and mechanical  properties control several processes in the cell and beyond. A key property of these filaments that allows them to be so flexible in their structure is their dynamic lability which allows them to grow or shrink, and become crosslinked or fall apart with relative ease. This is achieved through a variety of structural and chemical means such as capping, treadmilling and most spectacularly, dynamic instability. This last process which involves intermittent transitions between phases of slow growth  and rapid shrinkage of a polymer, first seen in microtubules  \cite{mitchison1984}. 

After the discovery of dynamic instability in microtubules \cite{mitchison1984}, various models \cite{flyvbjerg1994,antal2007,ranjith2009, brun2009} have been proposed to interpret the phenomenon as a stochastic process with different kinetic constants for the addition and removal of subunits from the ends of polar subunits. While this chemical kinetic approach leads to results that are able to explain the experimental observations qualitatively, over the years it has become increasingly clear that dynamic instability in microtubules, the best studied system to date, has an important structural component associated with the change in shape of the dimers once the attached GTP is hydrolyzed \cite{hyman1998,chretien1999,wang2005}. In particular, microtubules which are formed by a number of protofilaments, typically thirteen, grow by the addition of tubulin dimers which are in their GTP-bound state. Soon after polymerization, the GTP-bound tubulin changes conformation from a straight state to a bent state upon the hydrolysis of the GTP unit \cite{hyman1998,wang2005}. This conformational change is critical to dynamic instability, since curved filaments tend to detach from the microtubule while straight filaments are stable. Indeed electron micrographs of microtubules caught ''in flagrante delicto" show that individual protofilaments can be seen curving outwards from the frayed ends \cite{chretien1999}, and more recently individual protofilaments have been found to assemble into rings that curve along a direction orthogonal to that when they are part of the tubule \cite{wang2005}. Mechanical measurements of the rigidity of microtubles \cite{Kis} show that the Young's modulus of the assembly is two orders of magnitude smaller than its shear modulus, consistent with the structural evidence of strong interactions between tubulin dimers along a protofilament and weak lateral interactions between tubulin dimers on different protofilaments.  Taken together these observations suggest that the stability and dynamic instability of microtubules involves structural, mechanical and kinetic aspects \cite{janosi1998,janosi2002,mahadevan2005,hunyadi2007, review}.  Large scale computationally intensive models \cite{vanburen2005}  do try and to account for these effects, but  at the cost of understanding the generality and qualitative aspects of the basic phenomenon. 

Since the observations of dynamic instability in microtubules more than 20 years ago, the phenomenon has been implicated in the dynamics of single actin filaments \cite{fujiwara2002}, observed in the bacterial homolog of actin, ParM \cite{garner2004}, and is thought to also occur in bacterial homologs of microtubules \cite{popp2010}. In all these cases, the process of dynamic instability is characterized by the cooperative assembly and disassembly of a filamentous multi-stranded polymer that involves multiple structural states of the subunits and filament ends \cite{review} that are dependent on the hydrolysis of GTP or ATP. Thus, the lability of the polymeric filament is directly tied in to the kinetics of assembly as well as the consequent structural polymorphism, and conveniently used by the cell to direct its internal organization during movement, division and other functions. The appearance of similar kinetic profiles for the growth and shrinkage of polymeric filaments that are very different biochemically suggests a functionally-driven convergent evolution that selected certain traits in subunits capable of self assembly. This naturally raises the question of the minimal system that is capable of robust dynamic instability in polymeric assemblies  with the following ingredients (i) a slender geometry associated with both the individual protofilament and the filament assembly, (ii) bond interactions of different strengths along and across protofilaments,  (iii) kinetics of subunit addition being different from that of NTP hydrolysis in the added subunit. Each of these requirements is fairly generic and suggests a model of  a single elastic filament which can grow or shrink by subunit addition, can attach to or detach from a substrate (of  other protofilaments) and has in internal degree of freedom associated with an NTP-hydrolysis driven conformation change. 
Our model is similar in spirit to the filament model introduced in Ref. \cite{janosi2002}.

The relative simplicity of the model allows us to use a combination of scaling arguments, analytic solutions,  and  simple numerical simulations to characterize both the deterministic and stochastic aspects of the growing and shrinking phases of the elastic filament. In particular, we find that for the completely deterministic case, simple arguments allow us to characterize the mechanochemical conditions under which filaments either grow without bound or collapse and disappear. However, interestingly, in the presence of any disorder or fluctuations in the kinetic, structural or mechanical parameters, a window opens up between the growth and collapse phases that allows for the coexistence of both states, which naturally characterizes the region of dynamic instability in terms of experimentally measurable parameters. In addition to providing an explanation for the occurrence of dynamic instability in polymeric filaments,  our study may be of possible significance  for the evolvability of this trait.

\section*{Mathematical model}

For simplicity of exposition, we consider a one-dimensional elastic filament restricted to move in two dimensions and capable of attaching to or detaching from a rigid substrate with which the filament interacts via a series of springs (see Fig.~\ref{Fig1}a). The filament is assumed to be made of subunits of length $a$, and have a bending stiffness $B$ and stretching stiffness $E$ (where, as is usual $B/Ea^2 \ll 1$), while the linear springs  that may connect it to the substrate are assumed to have a stiffness $S$ and a maximum extension $r_c$. 

Given an initial state of the filament  with say $N$ segments that is attached to a substrate, we assume that its shape described by the positions of the ends of its segments ${\bf r_i}(t)= (x_i(t),y_i(t))$ evolves according to the equation 
\begin{equation}
\eta\dot{{\bf r}}_i= -\frac{\partial \cal E}{\partial {\bf r}_i} + \mathbf{f}_i(t) 
\label{eq:dyn}
\end{equation}
corresponding to overdamped Langevin dynamics with $\eta$ the damping coefficient, and the elastic energy of the filament $\cal E$ is given by
\begin{equation}
{\cal E}=\sum_{i=1}^N \frac{1}{2} E(|{\bf r}_{i+1}- {\bf r}_i|-a))^2-B \cos(\theta_i - \phi)+\frac{1}{2}S ({\bf r}_i-{\bf R}_i)^2,
\end{equation}
where $\theta_i$ are the angles between neighboring subunits,  and ${\bf R}_i=(a i,0)$ are the positions of the endpoints of the $i-$th spring on the substrate to which the filament is attached, with the proviso that $S = 0$, if $|{\bf r}_i-{\bf R}_i| \ge r_c$, i.e. the springs break when the extension equals or exceeds the threshold $r_c$. Here, 
the first term corresponds to the stretching energy (which is negligibly small relative to the other terms), the second term is the bending energy associated with rotating one segment relative to its neighbor  away from $\phi$ the natural rest angle between them, and the last term is the energy associated with the interactions with the substrate.
The thermal fluctuations are characterized by  $\mathbf{f}_i(t)$, with $\langle \mathbf{f}_i(t)\rangle=0$, and 
correlations
\begin{equation}
\langle f^{\gamma}_i(t)f^\delta_j(t') \rangle =2\eta k_B T \delta_{ij}\delta_{\gamma\delta}\delta(t-t'),
\end{equation}
where $f^\gamma_i$ is the  $\gamma$ component of the vector $\mathbf{f}_i$, $T$ is the temperature, $k_B$ is the Boltzmann constant, and $\delta_{ij}$ is the usual Kronecker-delta operator.The assumptions underlying the above dynamics are common in polymer physics: we ignore all long range hydrodynamic interactions, and furthermore assume that detailed balance and the fluctuation-dissipation theorem are valid at the scale of the filament.    

Subunits can attach to an end of the filament with a rate $1/\tau_G$, where $\tau_G$ is a characteristic growth time. We note that for polar filaments, the rate of attachment on either end of the filament will be different; here we dispense with this difference for simplicity, but can easily account for the new effects that it will lead to, including treadmilling. Newly attached subunits start out collinear with the filament, but evolve to acquire  an intrinsic curvature $\kappa = \phi/a$ (where $\phi$ is the intrinsic angle between segments) via a single step first-order kinetic process that models the conformation change associated with NTP hydrolysis. The associated evolution equation for the curvature reads
\begin{equation}
 \frac{d \kappa}{dt} = \frac{1}{\tau_\kappa} (\kappa_0-\kappa),
\label{eq:kappa}
\end{equation}
where $\kappa_0=\phi_0/a$ is the intrinsic curvature at equilibrium following hydrolysis and $\tau_\kappa$ is the characteristic timescale for the evolution of curvature. 

In addition to ambient thermal fluctuations, the structural, kinetic and mechanical properties of the filament in general could fluctuate in space due to various source of local heterogeneities in hydrolysis rates, or the adhesion to the substrate to name just two possibilities. This type of heterogeneity leads to quenched disorder (i.e. time
independent disorder) and we will consider a particular manifestation of it as exemplified by a random distribution of the adhesion spring constants $S_i$ that leads to a random distribution of spring toughness along the filament.  In particular we use the Gamma distribution given by
\begin{equation}
p(S)= \frac{k^k S^{k-1} \exp(-kS/\mu)}{\mu^k\Gamma(k)},
\label{eq:gamma}
\end{equation}
where $k$ is an integer and $\mu$ is the mean of the distribution (i.e. $\langle S \rangle=\mu$). With this choice, changing $k$ allows us to keep the mean of the distribution constant and vary its standard deviation $\sigma=\mu/k$, allowing us to tune the degree of disorder and quantify it using the coefficient of variation $C_v=\sigma/\mu=1/k$. Recent experiments using varying concentrations of non-hydrolyzable analogs of tubulin and their variants in microtubules are consistent with this type of quenched disorder, as we discuss later.

There are two natural dimensionless parameters in the problem: (i) $\alpha^2 = B\kappa^2/Sr_c^2$, a ratio of the linear energy density   associated with straightening out a filament with a natural curvature $\kappa$ and the linear energy density associated with the adhering springs at their point of failure $Sr_c^2$   and (ii) $\beta = \tau_\kappa/\tau_G$ a ratio of the time scales for hydrolysis to subunit addition, in addition to the coefficient of variation that characterizes the disorder $1/k$. As we will see these two parameters characterize the phase space of possibilities for the stability and dynamics of the filaments while the presence or absence of thermal and quenched disorder controls the appearance of a window of coexistence of the growing and shrinking phases where dynamic instability arises.  In Table \ref{table:par}, we list the main parameters employed in the model.

To study the dynamics of Eq.~\ref{eq:dyn}-Eq.~\ref{eq:kappa}, we prescribe an initial state of the filament with one end of the filament strongly attached to the substrate so that we need to track only the dynamics of the other end as it grows and shrinks. Our analysis starts with a consideration of the static problem before we progressively increase the level of complexity of the model. We start by considering the dynamics of the filament in the limit $\tau_G \rightarrow \infty$ (i.e. no growth), in a fully hydrolyzed state at temperature $T = 0$ and in absence of quenched disorder. We then separately study the effects of quenched disorder and thermal fluctuations. Finally, we consider the full model with growth and hydrolysis. In each case, we use a combination of analysis and numerical simulation; the latter is carried out using a conjugate gradient method (for the static case), and a fourth order adaptive step Runge-Kutta method in the limit $T = 0$ and a simpler Euler scheme in the case $T > 0$ (for the dynamic cases).

\section*{Analysis}
\subsection*{Statics: unbinding transition}
We start our analysis of the mathematical model by focusing on the simplest of cases corresponding to the static solution of Eq. \ref{eq:dyn} at $T=0$ (i.e. without noise)  which satisfies
\begin{equation}
\frac{\partial \cal E}{\partial {\bf r}_i}=0.
\label{eq:equil}
\end{equation}
In this setting, we let $\tau_G$ be larger than all other time scales in the problem, so that the filament has a constant length. Furthermore we impose that all the subunits have a constant intrinsic curvature $\kappa=\phi/a$. Then the system of equations Eq. \ref{eq:equil} can be efficiently solved using the method of conjugate gradients in a manner that is much faster than a dynamic simulation, an approach we use to simulate the model for $B=1$,$E=10^3$,$r_c=1$, $a=1$, $N=128$ where we note that $B/Ea^2 \ll 1$ so that polymer stretching is unimportant.  Then the behavior of the model only depends on the dimensionless ratio $\alpha$, via the static subunit angle $\phi = \kappa a$ (which is constant here) and the lateral stiffness $S$.  For small $\phi$ the filament is completely attached; once $\phi$ is increased past a critical value,  the filament unbinds from the substrate starting from a free edge.  The unbinding  transition is abrupt as can be seen computing the steady-state value of the spatially averaged local bending angle $\sum_j \theta_j/N $ (or analogously the scaled curvature)  as a function of the intrinsic bending angle $\phi$. At a critical value $\alpha_c$, itself a function of $S$, we see that the filaments unbinds globally, as shown in Fig.~\ref{Fig1}b.


In order to obtain an analytical understanding of the unbinding transition, we consider the continuum limit and restrict ourselves to small slopes and deformations, parametrizing the filament coordinates in terms of the horizontal component $x$: ${\bf r}_i \to (x,y(x))$, where $y(x)$ describes the filament
shape. By neglecting stretching of the polymer, the quadratic energy functional can be written as 
\begin{equation}
 {\cal E} = \int \frac{1}{2}(aB(y^{\prime\prime}-\kappa)^2+Sy^2) dx.
\end{equation}
The Euler-Lagrange equations associated with minimizing ${\cal E}$ lead to the equilibrium equation (Eq.~\ref{eq:equil})  which now read
\begin{equation}
B y^{\prime\prime\prime\prime}+S y =0, \label{cont}
\end{equation}
The associated boundary conditions are
\begin{equation}
 y^{\prime\prime}(0)=\kappa \mbox{\hspace{1cm}} By^{\prime\prime\prime}(0)=-Sy(0)a/2,\mbox{\hspace{1cm}} y(\infty), y^\prime(\infty) \to 0
\end{equation}
Here the first boundary condition imposes the fact that the filament has an intrinsic natural curvature $\kappa$, the second boundary condition accounts for the discrete nature of the adhering spings and arises by noting that the force from the lattice of springs can be rewritten as $-S a\sum_i\delta(x-i a) y$, while the last two simply state that a semi-infinite filament strongly attached at infinity is horizontal. Solving the boundary value problem Eq. (8-9) leads to a shape profile for the filament given by
\begin{equation}
y(x)=\frac{\kappa}{2q^2(1+aS/(4Bq^3)) } e^{-qx}\cos(x)-\frac{\kappa}{2q^2}e^{-qx}\sin(qx). \label{cont-profile}
\end{equation}
where $q=(S/B)^{1/4}/\sqrt{2}$ and characterizes a natural healing length $l_h=1/q =\sqrt{2}(B/S)^{1/4}$. Thus over scales large compared to $l_h$ from the free edge, the filament remains straight when bound to the substrate. This linear solution agrees very well with the numerical solution as shown in Fig~\ref{Fig1}c. Not accounting for the discrete nature of the adhering springs is tantamount to setting $a=0$, so that the solution Eq. \ref{cont-profile} reduces to $y(x)=\frac{\kappa}{2q^2} e^{-qx} (\cos(qx)-\sin(qx))$ which slightly underestimates the numerical solution,  a result similar to that obtained in \cite{janosi2002} for a more complex cylindrical geometry. The analytical solution Eq. \ref{cont-profile} allows  us to determine the condition for unbinding of the filament which would naturally start at the free edge when $y(0) = r_c$, i.e. the spring at the edge has been stretched to its maximum length and would thus break. In the absence of fluctuations and disorder, this would cause the entire filament to unzip. If $\kappa_c$ is the critical natural curvature above which the filament will spontaneously unbind, we find that the critical condition reads $y(0)=\kappa_c/2q^2(1+aS_0/(4Bq^3))=r_c$, which  can be recast in terms of the dimensionless parameter $\alpha=\sqrt{B\kappa_c^{2}/Sr_c^2}$ yielding a critical structural parameter $\alpha_c=(1+\frac{a}{l_h})$ which delineates the bound and unbound states. In the inset Fig.~\ref{Fig1}b, we show that the rescaled numerically obtained critical curvature $\alpha_c$ fits the simple theoretical expression above very well, and as the healing length $l_h/a \rightarrow \infty$, $\alpha_c=1$ as expected. To understand the dynamical process of unbinding, we now consider the effects of both deterministic and stochastic processes.

\subsection*{Deterministic dynamics}

When $\alpha>\alpha_c$, we numerically integrate  the equations of motion Eq. (\ref{eq:dyn}) and follow the evolution of the length of the detached part of the filament, which is still assumed to be of uniform length over the time scale of observation. The results, shown in Fig.~\ref{Fig2}a, suggest that the detached length grows diffusively in time. This behavior can be explained by a simple scaling argument that balances the viscous dissipation rate $P_v \simeq \eta (l/t)^2 l$, where $l$ is the length of the detached filament  with the elastic power that drives unbinding $P_e\simeq (B \kappa^2 - S r_c^2) l/t= B(\kappa^2-(\kappa_c)^2)l/t$, and yields $l \sim (B(\kappa^2-(\kappa_c)^2)t/\eta)^{1/2}$. The slowing down of unbinding with time arises because the ever lengthening unbound part takes a longer and longer time to move through the viscous environment; eventually once the unbound part has formed a circular ring, this diffusive behavior will likely be replaced by linear Stokesian dynamics, although we do not reach this limit in our simulations.  In contrast, if the subunits break off as soon as they unbind from the substrate, viscous dissipation is localized to a region near the dynamic detachment zone, and the viscous dissipation rate  $P_v\simeq \eta (l/t)^2 a$ so that now $l \sim B(\kappa^2-(\kappa_c)^2)t/a \eta$. To verify this relation, we use numerical simulations where we remove the subunit when it detaches from the substrate; the results shown in  Fig. \ref{Fig2}b confirm that we indeed capture this Stokesian limit as well.

\subsection*{Dynamics with quenched disorder}

When we introduce quenched disorder into the structural parameter $\alpha_c$ via the dependence of the substrate stiffness $S$ through its coefficient of variation $C_v$, unbinding  occurs stochastically. In Fig.~\ref{Fig3}a, we show the results of simulations for the steady-state value of the average local bending angle $\langle\sum_j \theta_j/N \rangle$ as a function of the intrinsic bond angle $\phi=a \kappa$, for different values of the coefficient of variation $C_v$,
keeping the average stiffness constant (in this example $\mu=0.05$). Since every calculation starts with a particular realization of the quenched disorder, our results are shown as averages over different realizations of the disorder. 

We note that disorder causes the unbinding of the filament to occur for values of $\phi$ that are larger than when disorder is absent ($C_v=0$). The seemingly counter-intuitive result that disorder makes the system stronger is immediately rationalized once we realize that this is only true because unbinding is always ruled by the strongest region, unlike material fracture or failure that is controlled by the weakest bond \cite{alava2009}.  For $N$ independent random variables $S_i$, distributed according to Eq.~\ref{eq:gamma}, the probability that the maximum is less than $S$ is equivalent to the probability that all the values $S_i$ are less than $S$. Hence, the cumulative distribution of the maximum over $N$ values is given by
\begin{equation}
P_N(S)=P(S)^N 
\label{eq:maxn}
\end{equation}
where $P(S)\equiv \int_0^S p(x)\;dx$ is the cumulative distribution associated
to $p(x)$. We can estimate the unbinding threshold considering an equivalent
filament attached with springs whose stiffness is given by the average $S_{max}$ of the distribution 
in Eq. \ref{eq:maxn}. The result, in dimensionless terms, reads as
\begin{equation}
\alpha_c \simeq  (1+ a/l_h)\sqrt{S_{max}/\mu}.
\end{equation}
Our dynamical simulations allow us to deduce this relationship numerically, and as shown in the inset of Fig.~\ref{Fig3}a, there is  good agreement with the  simple theory. It is also worth pointing out, that in the thermodynamic limit of large $N$, the asymptotic limit of Eq.~\ref{eq:maxn} is given by the Gumbel distribution:
\begin{equation}
P_{N}(S) \simeq \exp[-Ne^{-kS/\mu}]
\label{eq:max}
\end{equation}
Then the average value of the stiffness is given by  $\langle S_{max}\rangle \simeq \mu(\gamma + \log N)/k$, where $\gamma\simeq 0.57$ is the Euler constant, or in dimensionless terms, $\alpha_c \simeq (1+a/l_h)\sqrt{(\gamma+ \log N)/k}$.

\subsection*{Dynamics with thermal fluctuations}

In the presence of thermal fluctuations but without any quenched disorder,  the unbinding of the filament can be activated by noise even though the filament is nominally stable from a deterministic perspective. This is because, at sufficiently low temperatures, one can reasonably assume that unbinding always starts from the edge of the filament and proceeds by breaking the bonds sequentially \footnote{internal breakages are of course possible, but they result in self-limiting catastrophes since any internal bubble of detachment will lead to a weakening of driving stress.}. Then suggests that  the typical unbinding velocity follows the Arrhenius law  \cite{fluxcreep} 
\begin{equation}
v \propto \exp\left(-\frac{\Delta E}{k_BT}\right),
\label{eq:thermal}
\end{equation}
 where the energy barrier is given by $\Delta E \simeq \frac{1}{2}(Sr_c^2-B\kappa^2)l_h \sim \alpha_c-\alpha$. Our numerical simulations of Eq. \ref{eq:dyn} in a system with $\alpha = 0.5 \alpha_c$ and different values of the temperature $T$ confirms this picture, as shown in Fig.~\ref{Fig3}b.

%

\subsection*{Kinetics with thermal and quenched disorder}

We finally tie all the elements of the minimal model together to study the kinetics of an adsorbed polymorphic filament that can grow and change conformation as characterized by the two natural control parameters in the problem: $\beta=\tau_\kappa/\tau_G$ the ratio of the rate of conformation change (NTP hydrolysis)  to the rate of addition of subunits  and the maximum value reached by the structural/mechanical control parameter $\alpha_0$ (where the subscript now denotes the equilibrium value of the parameter, which is also its maximum, in contrast with its dynamical value $\alpha$).  A natural link between the two dimensionless parameters in the problem can be seen by rewriting  Eq.~\ref{eq:kappa}  as $\alpha$ as $d\alpha/dt = \tau_\kappa(\alpha_0-\alpha)$, where $\alpha_0\equiv \kappa_0\sqrt{B}/r_c\sqrt{\mu}$
and $\mu$ is the average of the spring stiffness $S$. 

In Fig.~\ref{Fig4} we show the evolution of the scaled filament length $L/a$ for different value of $\alpha_0$ and $\beta$, with springs of random stiffness, distributed according to Eq.~\ref{eq:gamma} with $k=4$ and $\mu=0.05$.  In our simulations, we see that for constant $\beta$ the filament grows for small $\alpha_0$ and collapses for large $\alpha_0$ (Fig.~\ref{Fig4}a). Similarly, for constant $\alpha_0$, the filament grows for small $\beta$ and collapses for large $\beta$ (Fig.~\ref{Fig4}b).  However, for intermediate ranges of $\alpha_0, \beta$, we observe a coexistence of the growth and collapse regimes, punctuated by intermittent {\it rescue} events. In this case, we see that the rescue events often occur at the same location, consistent with our earlier analysis which shows that a single tough bond is sufficient to stop the fracture from propagating. However, once a new tough bond is formed further along the growing filament, it serves to arrest the next catastrophe. The heterogeneous nature of the observed dynamic instability in our minimal model is related to the presence of quenched disorder in the mechanical parameter $\alpha_0$, since the kinetics are assumed to be completely deterministic.  These results are consistent with recent experiments that show that rescue events are directly correlated with the presence of GTP-bound subunits in a microtubule \cite{Dimitrov2008}, which serve to arrest catastrophes and rescue them repeatedly at the same location.

If instead, we use a  structurally homogeneous system with no quenched disorder but with stochastic hydrolysis events, then  Eq.~\ref{eq:kappa} describes the process only on average. Allowing each subunit to be hydrolyzed stochastically with rate $r_H=1/\tau_\kappa$, switching its intrinsic curvature from $\kappa=0$ to $\kappa=\kappa_0$ (or equivalently from $\alpha=0$ to $\alpha=\alpha_0$) leads to a crossover from growth to collapse as $\beta$ is increase, as shown in Fig.~\ref{Fig4}c, but with one important difference - in the absence of quenched structural disorder, catastrophes are much more likely to go all the way to collapse. Of course varying the the type of stochasticity in the hydrolysis will likely lead to  different types of intermittency. 

In Fig.~\ref{Fig5} we show a phase diagram showing  growth, collapse and  dynamic instability as a function of $\alpha_0$ and $\beta$ for the case of structural disorder and deterministic hydrolysis. The boundary between growth and collapse in the fully deterministic limit can be understood using simple considerations by noting that the filament will collapse when its curvature reaches the unbinding point before a sufficient amount of stabilizing subunits are attached to its end. The unbinding point is reached in a time $t_c$ given by $\alpha(t_c)=\alpha_c$ and we can thus write the condition for collapse as $t_c = n \tau_G$, where $n$ is a constant. Inserting this condition in the solution of Eq. \ref{eq:kappa}, we obtain an estimate for the phase boundary as
\begin{equation}
\alpha_0(\beta)= \frac{\alpha_c}{1-e^{-n/\beta}}, \label{pb}
\end{equation}
that is in qualitative agreement with the numerical results. Adding quenched disorder opens up a lenitcular region along this boundary where co-existence of the growth and collapsed phases leads to dynamic instability. In our analysis, we have assumed that the filament dynamics is much faster than the growth kinetics and the evolution of the structural parameter $\alpha$. If we relax this constraint, our model could also show a regime, defined ``third state'' in Ref. \cite{janosi2002} where catastrophes are interrupted by the attachment of new segments; this will be the subject of future work.

\section*{Discussion}

Inspired by the occurrence of metastability of a growing and shrinking phase in a variety of polymeric assemblies, we have studied a minimal model of a kinetically polymorphic elastic filament adsorbed on a soft  substrate that exhibits a window of dynamic instability. In particular, we find that the existence of this window depends on the existence of disorder in a phase space that controls the growth and shrinkage of the filament. We characterize the range of behavior of the filament in terms of two parameters, a structural-mechanical parameter that characterizes the stored energy in the residually strained assembly, and a kinetic parameter that characterizes the relative rate of growth to the rate of internal conformation change (associated with NTP hydrolysis). In the absence of any disorder, we find that the criterion for filament to remain bound or not is characterized by a structural parameter that measures the relative strength of intra-filament to filament-substrate bonds. This parameter has a natural extension to multi-filament polymeric assemblies such as microtubules, actin and their homologs in terms of intra-filament and inter-filament interactions which are known to play an important role in the mechanics and dynamics of these objects. Furthermore, we find that if the filament unbinds, it does so via either diffusive or Stokesian dynamics, depending on whether the filament does not or does disassemble as it unbinds. 

The presence of disorder, either quenched or thermal, changes the unbinding picture qualitatively. Quenched disorder in the structural-mechanical parameter effectively increases the unbinding strength, a result that  initially seems counter-intuitive given that disorder normally makes a sample weaker and thus more susceptible to fracture or peeling. Here, in this one-dimensional system, the effect is opposite due to the fact that stability is ruled by the strongest link. Thermal fluctuations on the other hand lead to subcritical unbinding, with a creep velocity that follows a simple Arrhenius law as expected from studies of other similar systems \cite{fluxcreep}.  Although our treatment of the kinetic parameter is rather simple, it is able to capture the competition between the rate of subunit addition and internal conformation change, a process implicated in the existence of a GTP cap; here we show that this leads to natural undulatory shape near the edge similar to that proposed earlier for more complex models \cite{janosi2002}.  When we consider the kinetics of conformation change in addition, we find that a window of dynamic instability opens up at the boundary that separates unlimited growth from collapse in the presence of quenched disorder. In this region, the growth and shrinkage phases co-exist and our  phase diagram for its existence points directly to the quantitative conditions for the existence of dynamic instability for our simple model. This pair of dimensionless measures which characterize  structural stability and the kinetics of assembly and conformation change in the presence of randomness may be generalized to more complex geometries such as sheets, tubes, helices and beyond and raise an interesting question with evolutionary implications:  what combination of geometry,  mechanics and chemistry can lead to the conditions for dynamic instability, and thus the flexibility to build structures that are functional, and yet not permanent ? Understanding the chemical constraints on the ranges of $\alpha_0$ and $\beta$  is clearly an important next step, and an artificial approach to this might well involve using a combination of known non-hydrolyzable analogs of GTP (NTP) along with the ability to track the locations \cite{zanic2009} associated with quenched structural disorder.

\section*{Acknowledgments} We thank H-Y Liang for discussions.


\clearpage

\begin{figure}
\centering
 \caption{Filament conformation - statics. a) Schematic representation of the system consisting of an elastic filament with a non-zero natural curvature adhered to a substrate via a series of springs. Here the filament is akin to a protofilament while the substrate represents the "bath" of the other filaments.  b) Static phase diagram showing the unbinding threshold in terms of the average local bending angle as a function of the intrinsic angle $\phi$ for different values of the stiffness $S$.  The inset shows the critical value of the dimensionless parameter $\alpha$, that characterizes the ratio of the filament and substrate energy, as a function of the ratio between the healing length $l_h$ and the discretization step $a$. The solid line is the theoretical result, in excellent agreement with the simulations, while the dashed line is the asymptotic  result for $a\to 0$. c) The characteristic shape of the attached filament in the neighborhood of the edge shows the healing length characterizing the balance between filament and substrate deformation. The simulation results are compared with our linearized continuum theory.
\label{Fig1}}
\end{figure}


\begin{figure}
\centering
\caption{Filament dynamics for different values of the intrinsic curvature $\kappa$ and spring stiffness $S$, corresponding to the unbound phase $\alpha >\alpha_c$. The filament unbinds dynamically - there are two possible mechanisms that determine the kinetics of the process. (a) Diffusive unbinding occurs when the filament detaches partially but remains intact. In this case, the figure shows  that the detached length increases as $l/a =V^*(t/\eta )^{1/2}$. The inset shows that the coefficient $V^*$, for different values of $S$, follows a single linear function when plotted  as a function of $(\kappa^2-\kappa_c^2)^{1/2}$.  (b) Stokesian unbinding occurs when the subunits break off from the filament once they are detached. The Figure shows that the detachmed  length grows as $l/a=V(t/\eta)/a$. The inset shows that the coefficient $V$ scales as is a linear function of $(\kappa^2-\kappa_c^2)$ as expected for small to moderate values of the parameters. \label{Fig2}}
\end{figure}


\begin{figure}
\centering
\caption{Filament dynamics in the presence of quenched disorder and thermal fluctuations. (a) When quenched
disorder is introduced in the spring stiffness, the unbinding threshold in Figure 1 is broadened due to the statistics of extremes as characterized in terms of the coefficient of variation $C_v = \sigma/\mu$, the ratio of the standard deviation to the mean of the spring stiffness $S$. Inset shows the critical value of the dimensionless parameter $\alpha\equiv \sqrt{B}\kappa/\sqrt{\mu}r_c$  as a function of the coefficient of variation of the disorder. The error bar represents the variance of $\alpha_c$ rather than the error on the mean. Both the mean and the variance  of $\alpha_c$ are seen to increase with increasing $C_v$ as predicted by extreme value statistics (line).  (b) Due to thermal fluctuations, the filament can unbind also for $\alpha<\alpha_c$. Here, we show the decrease of the filament length for $\alpha=0.5 \alpha_c$. The inset shows that the velocity in the initial stage satisfies the Arrhenius law. \label{Fig3}}
\end{figure}


\begin{figure}
\centering
\caption{Filament growth and collapse kinetics. a. Filament length as a function of time for a range of $\alpha_0$ the mechanical control parameter. b. 
Filament length as a function of time for a range of $\beta=\tau_\kappa/\tau_G$, the ratio of the time constant for the intrinsic curvature to equilibrate relative to 
the time constant for subunit addition,  the biochemical control parameter. c. Filament length when the hydrolysis is random, also shows the transition 
between growth and catastrophe, but with one important difference from the case when the toughness parameter $\alpha$ is random; shrinkage goes all the way to collapse in the absence of any structural mechanical inhomogeneities. \label{Fig4}}
\end{figure}


\begin{figure}
\caption{A phase diagram in $\alpha-\beta$ space summarizing the regimes of growth, collapse and dynamic instability, a window that appears in the presence of disorder either in the structural toughness of the filament-substrate bond, or the hydrolysis. The dashed line is the theoretical prediction corresponding to Eq.~\ref{pb}.
\label{Fig5}}
\end{figure}

\begin{center}
\begin{table}
\begin{tabular}{ll}
\hline
 Parameter & Description\\ 
\hline
{\bf Elastic constants:} & \\
\hline
$B$ & filament bending stiffness\\ 
$E$ & filament stretching stiffness\\ 
$S$ & spring  stretching stiffness \\
\hline
{\bf Characteristic lengthscales: } & \\
\hline
$r_c$ & spring maximum extension \\ 
$a$ & segment length \\ 
$L$ & length of attached portion of the filament\\
$l$ & length of detached portion of the filament \\
$l_h \equiv 1/q = \sqrt{2}(B/S)^{1/4}$ & healing length \\
\hline
{\bf Angles and curvatures:} & \\
\hline
$\kappa$ & intrinsic curvature \\ 
$\phi= a \kappa $ & intrinsic angle \\ 
$\theta$ & local filament angle \\
$\kappa_0$ & maximum value of $\kappa$ when $\kappa$ is evolving\\
$\kappa_c$ & unbinding threshold\\ 
\hline
{\bf Dynamics:} & \\
\hline
$ \eta$ & filament damping coefficient\\ 
$ \tau_\kappa$ & timescale of curvature evolution\\
$\tau_G$ & timescale of filament growth\\
\hline 
{\bf Disorder:} & \\
\hline
$\mu$ & average spring stretching stiffness $S$ \\
$\sigma$ & standard deviation of stiffness distribution \\
$C_v=\sigma/\mu$ & coefficient of variation  \\
$S_{max}$ & maximum value of $S$ along the filament\\
\hline 
{\bf Control parameters:} & \\
\hline
$\alpha \equiv \sqrt{B}\kappa/\sqrt{S}r_c$ & mechanical control parameter \\
$\alpha_0$ & maximum value of $\alpha$ when $\kappa$ is evolving \\
$\beta \equiv \tau_\kappa/\tau_G$ & kinetic control parameter\\
\hline
\end{tabular}
\caption{\label{table:par} A description of the parameters of the model.}
\end{table}
\end{center}

\setcounter{figure}{0}

\begin{figure}
\centering
\includegraphics[width=8cm]{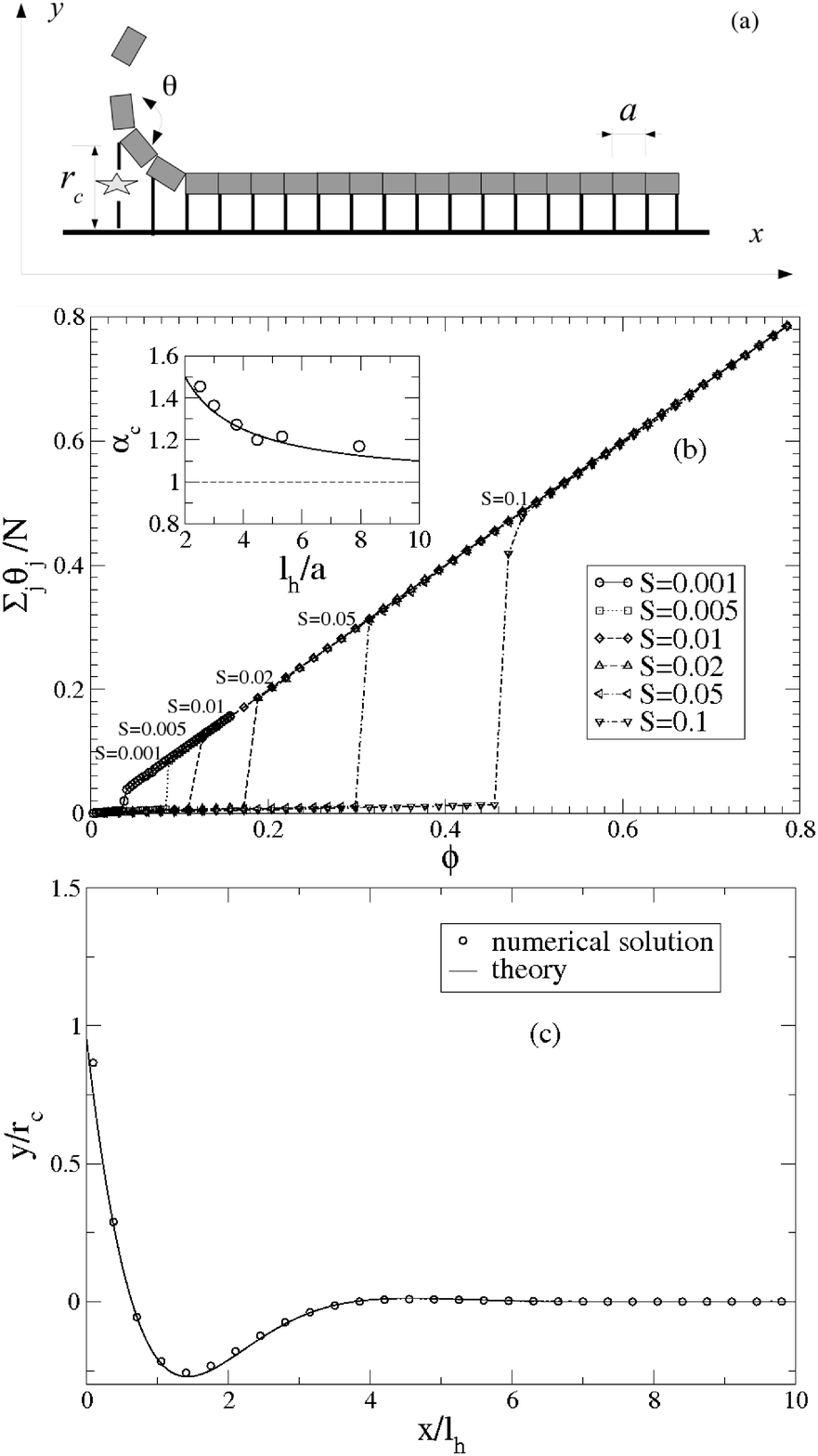}
\caption{}
\end{figure}

\clearpage

\begin{figure}
\centering
\includegraphics[width=8cm]{Fig2a_NEW.eps}

\vspace{1cm}

\includegraphics[width=8cm]{Fig2b_NEW.eps}
\caption{}
\end{figure}

\clearpage

\begin{figure}
\centering
\includegraphics[width=8cm]{Fig3a.eps}

\vspace{1cm}

\includegraphics[width=8cm]{Fig3b.eps}

\caption{}
\end{figure}

\clearpage

\begin{figure}
\centering
\includegraphics[width=8cm]{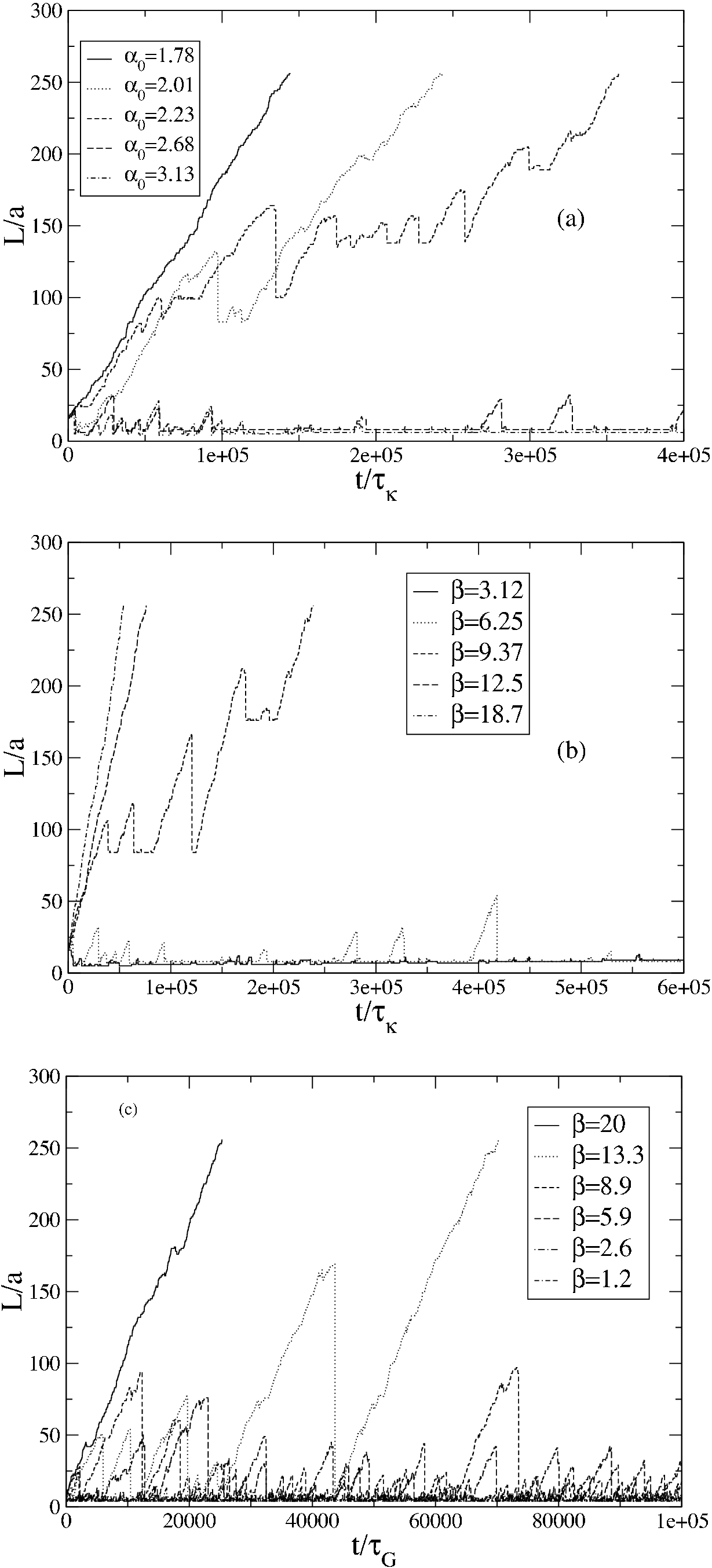}
\caption{}
\end{figure}

\clearpage

\begin{figure}
\includegraphics[width=7cm]{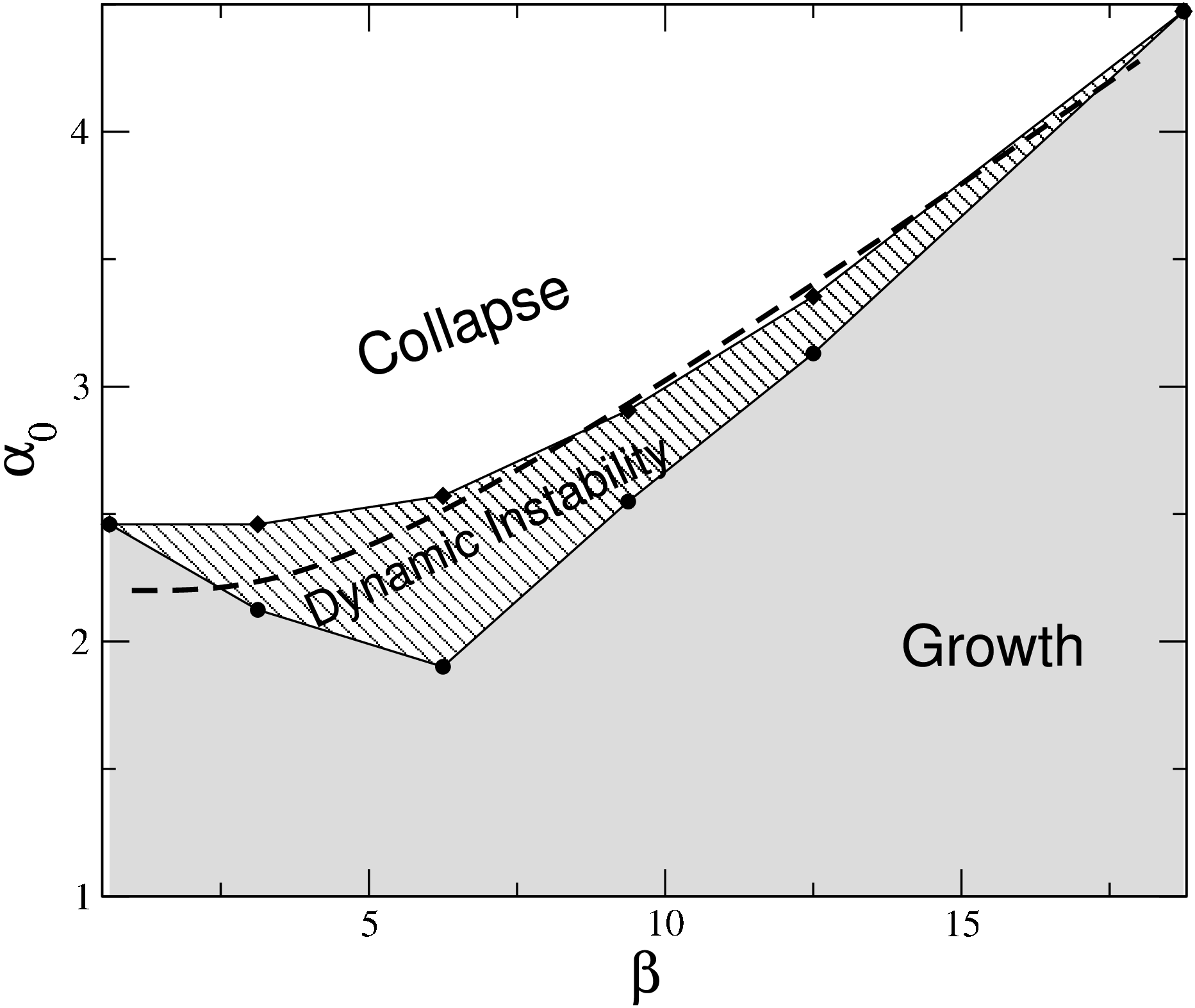}
\caption{}
\end{figure}

\end{document}